\documentclass[seceq]{iis}

\title{Variational Principle in Langevin Processes}
\runtitle{Variational Principle in Langevin Processes}

\author{%
\name{Yuki \surname{SUGHIYAMA}}$^{1}$
\CAE{yuki\_sughiyama@sat.t.u-tokyo.ac.jp} 
and \name{Masayuki \surname{OHZEKI}}$^{2}$ 
}

\inst{$^{1}$FIRST, Aihara Innovative Mathematical Modelling Project, Japan Science and Technology Agency, Institute of Industrial Science, The University of Tokyo, \address{Tokyo 153-8505, Japan}\\
$^{2}$Department of Systems Science, Graduate School of Informatics, Kyoto University, \address{Kyoto 606-8501, Japan}}

\runauthor{Yuki SUGHIYAMA and Masayuki OHZEKI}

\recdate{20YY}{0}{DD}
\accdate{20YY}{0}{DD}

\abst{The recent work, Nemoto and Sasa [Phys. Rev. E, {\bf 83}: 030105(R) (2011)], has shown that large deviations of the current characterizing a nonequilibrium system are obtained by observing the typical current for a modified system specified by a variational principle. In the present study, we will give a generalized version of the Nemoto-Sasa study by extracting a hidden mathematical structure from the fluctuation-response relation which is well-known in statistical mechanics. Here, the minimization of the Kullback-Leibler divergence plays an essential role.}

\kword{\kw{Nonequilibrium system}, \kw{Stochastic process}, \kw{Large deviation}}

\begin{document}

\maketitle

\section{Introduction}
As well-known in the statistical mechanics, the variance of the energy in equilibrium systems is evaluated by observing the specific heat, 
which means the linear response of the energy for the temperature change. 
Moreover, calculations of higher-order cumulants of the energy are performed on the basis of measurements for high-order responses. 
The formar is often called the fluctuation-response relation \cite{02}, which is one of epoch-making discoveries in statistical mechanics. 
Here, we widely use the term for the latter.
Such relationships exist not only between the energy and the temperature change but also between the magnetization and the susceptibility.

To be more precise, we discuss the fluctuation-response relation in terms of the large deviation theory \cite{03,04,05}, which deals with the large fluctuation i.e. higher-order cumulants. 
When we prepare the environment surrounding a system (i.e. the temperature, the magnetic field and so on are given), the energy density of the system is uniquely determined in the thermodynamic limit. 
We call such an energy density the internal or the typical energy density. 
Then, the large fluctuation around the typical energy density is expressed by the large deviation function defined as 
\begin{equation}
I_{\beta}\displaystyle \left(\epsilon\right)\equiv\lim_{N\rightarrow\infty}-\frac{1}{N}\log P_{N,\beta}\left(\epsilon\right),\label{ysm1}
\end{equation}
where $\epsilon$ and $N$ denote the energy density and the system size, respectively, and $P_{N,\beta}\left(\epsilon\right)$ represents the canonical density with the inverse temperature $\beta$. 
Here, the typical energy density $\epsilon_{*}\left(\beta\right)$ is given by $I_{\beta}\left(\epsilon_{*}\right)=0$. 
Under this setup, the fluctuation-response relation claims that the Legendre transform of the large deviation function (\ref{ysm1}), which is called the scaled cumulant generating function, can be evaluated 
by the typical energy density of the system with a different temperature (that is $\epsilon_{*}\left(\beta^{\prime}\right)$ ) [see section 2]. This function is closely related to the free-energy density. 

In this study, by extracting a hidden mathematical structure from the fluctuation-response relation, 
we attempt to formulate the method to evaluate the large deviation of the statistic (e.g. the sample mean and the empirical density) characterizing the original system 
by observing typical quantities of appropriately modified systems. 
Here, we show that modified systems are naturally given by a variational principle described by employing the Kullback-Leibler divergence. 
As a result, our formulation can be applied to not only equilibrium systems expressed by the canonical density but also noequilibrium systems described by Langevin processes. 

\section{Fluctuation-response relation}
To be self-contained, this section is devoted to briefly review the derivation of the fluctuation-response relation between the fluctuation of the energy density and the response for the temperature change. 
Suppose that a system consists of $N$ degrees of freedom represented by $\Sigma=\left\{\sigma_{1},\sigma_{2},\cdots,\sigma_{N}\right\}$ and is in an equilibrium state with an inverse temperature $\beta$.
The probability density of the system is described by the canonical density,
\begin{equation}
P_{\mathrm{e}\mathrm{q},\beta}\displaystyle \left(\Sigma\right)=\frac{1}{Z_{\beta}}e^{-\beta H\left(\Sigma\right)},
\end{equation}
where $H$ represents the Hamiltonian of the system and $Z_{\beta}$ denotes the partition function with the inverse temperature $\beta$. 
Under this setup, we calculate the scaled cumulant generating function of the energy density $\epsilon\equiv H/N$,
\begin{equation}
\displaystyle \lambda_{\beta}\left(k\right)=\lim_{N\rightarrow\infty}\frac{1}{N}\log\left\langle e^{Nk\epsilon\left(\Sigma\right)}\right\rangle_{\mathrm{e}\mathrm{q},\beta},
\end{equation}
where  the braket $\left\langle\cdot\right\rangle_{\mathrm{e}\mathrm{q},\beta}$ represents the expectation with respect to the canonical density $P_{\mathrm{e}\mathrm{q},\beta}$. 
By using the free energy density $f\displaystyle \left(\beta\right)=-\beta^{-1}\lim_{N\rightarrow\infty}\left(1/N\right)\log Z_{\beta}$, we obtain
\begin{equation}
\lambda_{\beta}\left(k\right)=\beta f\left(\beta\right)-\left(\beta-k\right)f\left(\beta-k\right).\label{ii1}
\end{equation}
The differentiation of both sides in Eq. (\ref{ii1}) leads us to  
\begin{equation}
\displaystyle \frac{\partial\lambda_{\beta}\left(k\right)}{\partial k}=\epsilon_{*}\left(\beta-k\right),\label{ii2}
\end{equation}
where we employ the thermodynamic relation, $\partial\left\{\beta f\left(\beta\right)\right\}/\partial\beta=\epsilon_{*}\left(\beta\right)$. 
Owing to Eq. (\ref{ii2}), we can evaluate the scaled cumulant generating function, which have all information for the fluctuation of the system, by observing the internal energy density of the system whose temperature is shifted by $k$. 
Accordingly, this relation (\ref{ii2}) is called the fluctuation-response relation.
By differentiating both sides in Eq. (\ref{ii2}) and setting $k$ as zero, we find the well-known form of the fluctuation-response relation, which is a relationship between the variance of the energy density and the specific heat as
\begin{equation}
N\left(\left\langle\epsilon^{2}\right\rangle_{\mathrm{e}\mathrm{q},\beta}-\left\langle\epsilon\right\rangle_{\mathrm{e}\mathrm{q},\beta}^{2}\right)=\beta^{-2}C\left(\beta\right),
\end{equation}
where $C\left(\beta\right)$ denotes the specific heat of the equilibrium state with the inverse temperature $\beta$.
Here, we used 
\begin{eqnarray}
\displaystyle \left.\frac{\partial^{2}\lambda_{\beta}\left(k\right)}{\partial k^{2}}\right|_{k=0}&=&N\displaystyle \left(\left\langle\epsilon^{2}\right\rangle_{\mathrm{e}\mathrm{q},\beta}-\left\langle\epsilon\right\rangle_{\mathrm{e}\mathrm{q},\beta}^{2}\right),\\
\displaystyle \left.\frac{\partial\epsilon_{*}\left(\beta-k\right)}{\partial k}\right|_{k=0}&=&\displaystyle \beta^{-2}C\left(\beta\right).
\end{eqnarray}

\section{General framework}
By extracting a mathematical structure from the fluctuation-response relation, 
we construct the general framework to evaluate the large deviation function from typical quantities of appropriately modified systems. Here, a variational principle plays a essential role. 
Let $X=\left\{x_{1},x_{2},\cdots,x_{N}\right\}$ be a sequence of random variables, and its probability density is given as $P_{0}\left(X\right)$. 
The macroscopic variable (e.g. the energy density, the magnetization and so on) is represented as $S\left(X\right)$. 
Then, we assume that $S$ satisfies the law of large number, that is $S$ almost surely converges to a typical value $s_{*}\left[P_{0}\right]=\left\langle S\left(X\right)\right\rangle_{0}$ in the limit of $ N\rightarrow\infty$, 
where the braket $\left\langle\cdot\right\rangle_{0}$ represents the expectation with respect to $P_{0}$. 
In order to calculate the large deviation function of $S$ in the probability density $P_{0}\left(X\right)$, we define the probability density of modified systems as $P_{k}\left(X\right)$. Here, $k$ denotes a modification parameter, and $P_{k}$ is given by 
\begin{equation}
P_{k}\displaystyle \left(X\right)\equiv\frac{e^{NkS\left(X\right)}}{\left\langle e^{NkS\left(X\right)}\right\rangle_{0}}P_{0}\left(X\right).\label{fluctuation-modified}
\end{equation}
We then obtain the following equation.
\begin{eqnarray}
\nonumber s_{*}\left[P_{k}\right]&=&\left\langle S\left(X\right)\right\rangle_{k}\\
\displaystyle \nonumber&=&\displaystyle \frac{1}{\left\langle e^{NkS\left(X\right)}\right\rangle_{0}}\int S\left(X\right)e^{NkS\left(X\right)}P_{0}\left(X\right)\prod_{i}dx_{i}\\
\displaystyle \nonumber&=&\displaystyle \frac{\partial}{\partial k}\left(\frac{1}{N}\log\left\langle e^{NkS\left(X\right)}\right\rangle_{0}\right)\\
&=&\displaystyle \frac{\partial\lambda_{0}\left(k\right)}{\partial k},\label{expectation-cumulant}
\end{eqnarray}
where $\lambda_{0}$ represents the scaled cumulant generating function of the original probability density $P_{0}$.
Comparing Eq. (\ref{expectation-cumulant}) to Eq. (\ref{ii2}), we find that Eq. (\ref{expectation-cumulant}) plays the same role as the fluctuation-response relation.
That is, we can evaluate the scaled cumulant generating function corresponding to $P_{0}$ by observing the typical value of $S$ in modified systems described by $P_{k}$.

Now, let us show an alternative way to give the modified probability density $P_{k}\left(X\right)$. 
Consider to maximize the following functional. 
\begin{equation}
\displaystyle \max_{P^{\prime}}\biggl\{k\int S\left(X\right)P^{\prime}\left(X\right)\prod_{i}dx_{i}-D_{\mathrm{K}\mathrm{L}}\left[P^{\prime}|P_{0}\right]\biggr\},\label{ysm3}
\end{equation}
where $D_{\mathrm{K}\mathrm{L}}$ denotes the scaled Kullback-Leibler divergence between $P_{0}$ and $P^{\prime}$, that is 
\begin{equation}
D_{\mathrm{K}\mathrm{L}}\displaystyle \left[P^{\prime}|P_{0}\right]\equiv\frac{1}{N}\int P^{\prime}\log\left(\frac{P^{\prime}}{P_{0}}\right)\prod_{i}dx_{i}.
\end{equation}
By employing the Lagrange multiplier method, the maximum of the funcrional (\ref{ysm3}) is given by 
\begin{eqnarray}
\displaystyle \nonumber 0&=&\displaystyle \frac{\delta}{\delta P^{\prime}\left(X\right)}\biggl\{k\int S\left(X\right)P^{\prime}\left(X\right)\prod_{i}dx_{i}-D_{\mathrm{K}\mathrm{L}}\left[P^{\prime}|P_{0}\right]-\alpha\left(\int\prod_{i}dx_{i}P^{\prime}\left(X\right)-1\right)\biggr\}\\
&=&\displaystyle \frac{1}{N}\log\frac{P^{\prime}\left(X\right)}{P_{0}\left(X\right)}-kS\left(X\right)-\left(\alpha-\frac{1}{N}\right),\label{ysm2}
\end{eqnarray}
where $\alpha$ denotes the Lagrange multiplier determined by the normalization condition for $P^{\prime}$,
\begin{equation}
\displaystyle \int P^{\prime}\left(X\right)\prod_{i}dx_{i}=1.
\end{equation}
By solving Eq. (\ref{ysm2}), we obtain 
\begin{eqnarray}
\displaystyle \nonumber \arg\max_{P^{\prime}}\biggl\{k\int S\left(X\right)P^{\prime}\left(X\right)\prod_{i}dx_{i}-D_{\mathrm{K}\mathrm{L}}\left[P^{\prime}|P_{0}\right]\biggr\}&=&P_{0}\displaystyle \left(X\right)e^{NkS\left(X\right)}e^{N\alpha-1}\\
&=&P_{0}\displaystyle \left(X\right)\frac{e^{NkS\left(X\right)}}{\left\langle e^{NkS\left(X\right)}\right\rangle_{0}}.\label{ysm4}
\end{eqnarray}
Comparing Eq. (\ref{fluctuation-modified}) to the right-hand side in Eq. (\ref{ysm4}), we find that the modified probability density $P_{k}\left(X\right)$ is given by the variational form, 
\begin{equation}
P_{k}\displaystyle \left(X\right)=\arg\max_{P^{\prime}}\biggl\{k\int S\left(X\right)P^{\prime}\left(X\right)\prod_{i}dx_{i}-D_{\mathrm{K}\mathrm{L}}\left[P^{\prime}|P_{0}\right]\biggr\}.\label{pk}
\end{equation}

Let us evaluate the scaled cumulant generating function, that is equivalent to the large deviation function, by employing this variational form. 
Calculating the maximum of the functional (\ref{ysm3}) by use of Eq. (\ref{ysm4}), we obtain
\begin{eqnarray}
\displaystyle \nonumber\max_{P^{\prime}}\biggl\{k\int S\left(X\right)P^{\prime}\left(X\right)\prod_{i}dx_{i}-D_{\mathrm{K}\mathrm{L}}\left[P^{\prime}|P_{0}\right]\biggr\}&=&k\displaystyle \left\langle S\left(X\right)\right\rangle_{k}-\frac{1}{N}\int\prod_{i}dx_{i}P_{k}\left(X\right)\log\frac{e^{NkS\left(X\right)}}{\left\langle e^{NkS\left(X\right)}\right\rangle_{0}}\\
&=&k\displaystyle \left\langle S\left(X\right)\right\rangle_{k}-k\left\langle S\left(X\right)\right\rangle_{k}+\frac{1}{N}\log\left\langle e^{NkS\left(X\right)}\right\rangle_{0}.
\end{eqnarray}
As a consequence, we reach
\begin{equation}
\displaystyle \lambda_{0}\left(k\right)=\max_{P^{\prime}}\biggl\{k\int S\left(X\right)P^{\prime}\left(X\right)\prod_{i}dx_{i}-D_{\mathrm{K}\mathrm{L}}\left[P^{\prime}|P_{0}\right]\biggr\}.\label{ysm5}
\end{equation}
These two equations (\ref{pk}) and (\ref{ysm5}) provide the method to evaluate the large deviation function by using the variational principle.
The reason is as follows. 
The diffrentiation of both sides in Eq. (\ref{ysm5}) leads us to 
\begin{equation}
\displaystyle \frac{\partial\lambda_{0}\left(k\right)}{\partial k}=\left\langle S\left(X\right)\right\rangle_{k}=s_{*}\left[P_{k}\right].\label{gfrr}
\end{equation}
Thus, we can calculate the scaled cumulant generating function by observing a typical value of $S$ in modified systems $P_{k}$, 
which is specified by Eq. (\ref{pk}). 
From the G\"{a}rtner-Ellis theorem \cite{05,26,27}, 
the Legendre transform of the scaled cumulant generating function gives us the large deviation function.
If we choose the original probability density $P_{0}$ as the canonical density describing equilibrium systems and the macroscopic variable $S$ 
as the energy density, Eq. (\ref{gfrr}) reduces to the ordinary fluctuation-response relation Eq. (\ref{ii2}). 
In the following section, we will apply this framework to Langevin systems.

Before closing this section, we attempt to express the large deviation function by using the scaled Kullback-Leibler divergence. 
We divide the maximizing procedure in Eq. (\ref{ysm5}) with respect to $P^{\prime}$ into two steps.  First, we minimize the Kullback-Leibler divergence under the constraint for the expectation, 
\begin{equation}
\displaystyle \int S\left(X\right)P^{\prime}\left(X\right)\prod_{i}dx_{i}=s.\label{const}
\end{equation}
Second, we maximize the functional by $s$, which is the expectation of $S\left(X\right)$ with respect to $P^{\prime}$.
That is, we write the maximization in Eq. (\ref{ysm5}) as 
\begin{equation}
\displaystyle \max_{P^{\prime}}\biggl\{k\int S\left(X\right)P^{\prime}\left(X\right)\prod_{i}dx_{i}-D_{\mathrm{K}\mathrm{L}}\left[P^{\prime}|P_{0}\right]\biggr\}=\max_{s}\left\{ks-\min_{P^{\prime}:(\ref{const})}D_{\mathrm{K}\mathrm{L}}\left[P^{\prime}|P_{0}\right]\right\},
\end{equation}
where $\displaystyle \min_{P^{\prime}:(\ref{const})}$ stands for the minimization with respect to $P^{\prime}$ under the constraint (\ref{const}). 
Accordingly, we obtain
\begin{eqnarray}
\displaystyle \lambda_{0}\left(k\right)=\max_{s}\left\{ks-\min_{P^{\prime}:(\ref{const})}D_{\mathrm{K}\mathrm{L}}\left[P^{\prime}|P_{0}\right]\right\}.\label{gnst}
\end{eqnarray}
Taking into account the fact that Eq. (\ref{gnst}) is written in the form of the Legendre transform, we simply obtain the large deviation function as 
\begin{equation}
I_{0}\displaystyle \left(s\right)=\min_{P^{\prime}:(\ref{const})}D_{\mathrm{K}\mathrm{L}}\left[P^{\prime}|P_{0}\right].\label{rate-kl}
\end{equation}

\section{Large deviation function for Langevin processes}
In order to confirm the generality of our framework discussed in the previous section, we show its applicability. 
We evaluate the large deviation function for the current of Brownian particles. 
This work has already been performed by Nemoto and Sasa \cite{22,23} in terms of eigenvalue problems. 
Here, we will find that the Nemoto-Sasa study is included as a part of our general framework. 
Since Brownian motions are often described by Langevin processes, we suppose 
\begin{equation}
dx_{t}=A\left(x_{t}\right)dt+B\circ dW,\label{ole}
\end{equation}
where $x$ represents the position of the Brownian particle with the periodic boundary condition (i.e. $x=x+L$), 
$W$ denotes the Wiener process \cite{28}  (i.e. the Gaussian white noise), and the symbol $\circ$ stands for the multiplication in the sense of the Stratonovich \cite{28,29}. 
Here, $B$ implies the noise intensity related to the temperature of the environment, and $A\left(x\right)$ is a deterministic force given as $ A\left(x\right)=-dV\left(x\right)/dx+\gamma$.
Due to the term $\gamma$, Eq. (\ref{ole}) describes the nonequilibrium system with the stationary current. 
Let us apply our framework to this system. 
Suppose the sequence of random variables $X$ as the realization of the Langevin equation (\ref{ole}). Then, the original probability density $P_{0}\left(X\right)$ mentioned in the variational priciple, Eqs. (\ref{pk}) and (\ref{ysm4}), is given as the conditional path probability density 
\begin{equation}
P_{0}\displaystyle \left(X|x_{0}\right)\propto\exp\left\{-\frac{1}{2B^{2}}\int_{0}^{\tau}dt\left[\left(\dot{x}-A\right)^{2}+B^{2}\frac{dA}{dx}\right]\right\},\label{pdo}
\end{equation}
where $x_{0}$ denotes the initial condition of the realization $X$, and $\tau$ is the time interval of measurements.
Here, Eq. (\ref{pdo}) is calculated from Eq. (\ref{ole}) \cite{28,29}.  
According to the variational principle (\ref{pk}), we seek the modified probability density $P_{k}\left(X\right)$ by tuning the probability density $P^{\prime}\left(X\right)$. Now, we attempt to perform this procedure by controlling an additional force $u\left(x\right)$. That is, we assume that the tuning probability density $P^{\prime}\left(X\right)$ is generated by the modified process, 
\begin{equation}
dx_{t}=\left\{A\left(x_{t}\right)+u\left(x_{t}\right)\right\}dt+B\circ dW,\label{mle}
\end{equation}
where $u\left(x\right)$ denotes the additional force for tuning $P^{\prime}\left(X\right)$ and is controllable function by external clients. 
One may say that we cannot find the optimal solution of the variational principle (\ref{pk}) by using this procedure, since arbitrary path probability density on $X$ cannot be expressed by tuning $u\left(x\right)$. 
However, we can show that the optimal probability density, which is the modified density $P_{k}$, is given by tuning $u\left(x\right)$ [detailed discussions are written in the Appendix A]. 
By using Eq. (\ref{mle}), the tuning probability density $P^{\prime}\left(X\right)$ is give by the conditional path probability density, 
\begin{equation}
P_{u}\displaystyle \left(X|x_{0}\right)\propto\exp\biggl\{-\frac{1}{2B^{2}}\int_{0}^{\tau}dt\left[\left\{\dot{x}-\left(A\left(x\right)+u\left(x\right)\right)\right\}^{2}+B^{2}\frac{d\left(A+u\right)}{dx}\right]\biggr\}.\label{pdm}
\end{equation}
By substituting Eqs. (\ref{pdo}) and (\ref{pdm}) into Eq. (\ref{ysm5}) and 
taking into account the fact that modified processes are parameterized by $u\left(x\right)$, 
we obtain the scaled cumulant generating function in the limit of $\tau\rightarrow\infty$ as
\begin{equation}
\displaystyle \lambda_{0}\left(k\right)=\max_{u}\left\{k\int\mathfrak{D}XS\left(X\right)P_{u}\left(X|x_{0}\right)-D_{\mathrm{K}\mathrm{L}}\left[P_{u}|P_{0}\right]\right\},\label{ucumulant}
\end{equation}
where $\int \mathfrak{D}X$ denotes integration over all realizations, and we use the fact that $\tau$ plays a role of $N$ in the previous section. 
Now, suppose that $S$ be a time-averaged current, $S\left(X\right)=\left(1/\tau\right) \int_{0}^{\tau} \dot{x}dt$. 
Then, we can evaluate the first term in the right-hand side in Eq. (\ref{ucumulant}) as 
\begin{eqnarray}
k\displaystyle \int\mathfrak{D}XS\left(X\right)P_{u}\left(X|x_{0}\right)=k\int dxJ_{\mathrm{s}\mathrm{t},u}\left(x\right),\label{jm}
\end{eqnarray}
where $J_{\mathrm{s}\mathrm{t},u}$ denotes the statinary current density in the modified process and is given by 
\begin{equation}
J_{\mathrm{s}\mathrm{t},u}\displaystyle \left(x\right)=\left\{A\left(x\right)+u\left(x\right)\right\}\rho_{\mathrm{s}\mathrm{t},u}\left(x\right)-\frac{B^{2}}{2}\frac{\partial\rho_{\mathrm{s}\mathrm{t},u}\left(x\right)}{\partial x}.
\end{equation}
Here, $\rho_{\mathrm{s}\mathrm{t},u}$ is the stationary probability density corresponding to Eq. (\ref{mle}).
On the other hand, by use of Eqs. (\ref{pdo}) and (\ref{pdm}), the second term in the right-hand side in Eq. (\ref{ucumulant}) is evaluated as
\begin{eqnarray}
\nonumber D_{\mathrm{K}\mathrm{L}}\left[P_{u}|P_{0}\right]&=&\left\langle\frac{1}{\tau}\log\frac{P_{u}\left(X|x_{0}\right)}{P_{0}\left(X|x_{0}\right)}\right\rangle_{u}\\
&=&\displaystyle \int dx\rho_{\mathrm{s}\mathrm{t},u}\left(x\right)\frac{u^{2}\left(x\right)}{2B^{2}},\label{klm}
\end{eqnarray}
where $\left\langle\cdot\right\rangle_{u}$ denotes the expectation with respect to $P_{u}\left(X|x_{0}\right)$ [detailed calculations are written in Appendix B].
By substituting Eqs. (\ref{jm}) and (\ref{klm}) into Eq. (\ref{ucumulant}), we find
\begin{equation}
\displaystyle \lambda_{0}\left(k\right)=\max_{u}\left\{\int dx\left[kJ_{\mathrm{s}\mathrm{t},u}\left(x\right)-\rho_{\mathrm{s}\mathrm{t},u}\left(x\right)\frac{u^{2}\left(x\right)}{2B^{2}}\right]\right\}.\label{ns1}
\end{equation}
The differetiation of both sides in Eq. (\ref{ns1}) with $k$ leads us to the form of the fluctuation-response relation, 
\begin{equation}
\displaystyle \frac{\partial\lambda_{0}}{\partial k}=\left\langle\frac{1}{\tau}\int_{0}^{\tau}\dot{x}dt\right\rangle_{u_{k}}.\label{gfrr2}
\end{equation}
Here, the additional force to specify the large deviation function for the original process, $u_{k}\left(x\right)$, is determined by
\begin{equation}
u_{k}\displaystyle \left(x\right)=\arg\max_{u}\left\{\int dx\left[kJ_{\mathrm{s}\mathrm{t},u}\left(x\right)-\rho_{\mathrm{s}\mathrm{t},u}\left(x\right)\frac{u^{2}\left(x\right)}{2B^{2}}\right]\right\}.\label{ns2}
\end{equation}
These equations ensure that we can evaluate the large deviation function for the current of the Brownian particle. 
One can find that Eqs. (\ref{ns1}), (\ref{gfrr2}) and (\ref{ns2}) are similar to Eqs. (\ref{pk}), (\ref{ysm5}) and (\ref{gfrr}). Accordingly, we obtain the following prescription. 
That is, (I) we construct modified stochastic processes by controlling the additional force following the variational principle Eq. (\ref{ns2}); 
(II) we observe the typical time-averaged current for each modified process. 
Expressons of Eqs. (\ref{ns1}) and (\ref{ns2}) reproduce the result in the Nemoto-Sasa study. 

\section{Summary}
In this study, we have formulated the method to evaluate the large deviation function of macroscopic variables in general stochastic systems by measurement for the typical quantity of modified systems. 
Here, we have shown that modified systems are given by the variational principle (\ref{pk}), 
which is written by the scaled Kullback-Leibler divergence. 
Applying our framework to the evaluation of the large deviation function for the current on Langevin processes 
describing the Brownian motion, we have found the Nemoto-Sasa study. 

From the discussion in sections 3 and 4, we find that our framework need not restrict the macroscopic quantity $S$ to the current. 
Therefore, we can evaluate the large deviation for arbitrary statistcs on Langevin processes. 
In particular, one may be interested in the evaluation of the large deviation for the occupation density, 
which is $\mu\left(y\right)\equiv\left(1/\tau\right) \int_{0}^{\tau} \delta\left(y-x_{t}\right)dt$, where $x_{t}$ is the solution of Eq. (\ref{ole}). 
When we apply our framework to this case, it is expected that we obtain the so-called Donsker-Varadhan formula \cite{03,04,05,24},
\begin{equation}
I_{0}\displaystyle \left[\mu\right]=-\min_{\phi>0}\left\{\int dx\mu\left(x\right)\frac{\hat{L}^{\dagger}\phi\left(x\right)}{\phi\left(x\right)}\right\},\label{dv}
\end{equation}
where $\hat{L}^{\dagger}$ denotes the adjoint of the Fokker-Planck operator \cite{28,29}, and $\phi$ is the variation parameter.
This study will be discussed in our sequel paper \cite{99}. 

\section*{Acknowledgments}
Y. S. would like to thank Prof. Yoichiro Takahashi for fruitful discussions on the large deviation theory. 
M. O. thanks financial support by MEXT in Japan, Grant-in-Aid for Young Scientists (B) No. 24740263.
This research is supported by the Aihara Innovative Mathematical Modelling Project, the Japan Society for the Promotion of Science (JSPS) 
through the "Funding Program for World-Leading Innovative R ampersand D on Science and Technology (FIRST Program)," initiated by the Council for Science and Technology Policy (CSTP). 

\appendix
\section{Optimal solution $P_{k}$ by tuning the additional force $u$}
We show that the optimal solution of the variational principle (\ref{pk}) is attained by tuning the additional force $u\left(x\right)$. 
The optimal solution $P_{k}\left(X|x_{0}\right)$ is written by 
\begin{eqnarray}
\displaystyle \nonumber P_{k}\left(X|x_{0}\right)&=&\displaystyle \arg\max_{P^{\prime}}\biggl\{k\int \mathfrak{D}XS\left(X\right)P^{\prime}\left(X|x_{0}\right)-D_{\mathrm{K}\mathrm{L}}\left[P^{\prime}|P_{0}\right]\biggr\}\\
&=&\displaystyle \frac{e^{\tau kS\left(X\right)}}{\left\langle e^{\tau kS\left(X\right)}\right\rangle_{0}}P_{0}\left(X|x_{0}\right),\label{add1}
\end{eqnarray}
where we recall Eqs. (\ref{fluctuation-modified}) and (\ref{pk}), and $S\left(X\right)$ is the time-averaged current, $S\left(X\right)=\left(1/\tau\right) \int_{0}^{\tau} \dot{x}dt$. 
Substituting the form of the time-averaged current into Eq. (\ref{add1}), we obtain
\begin{equation}
P_{k}\left(X|x_{0}\right)=\exp\left[k\int_{0}^{\tau}dt\left\{\dot{x}-\lambda_{0}\left(k\right)\right\}\right]P_{0}\left(X|x_{0}\right),\label{add2}
\end{equation}
where $\lambda_{0}\left(k\right)$ denotes the scaled cumulant generating function. 
On the other hand, the tuning probability density is given by Eq. (\ref{pdm}). That is
\begin{equation}
P_{u}\displaystyle \left(X|x_{0}\right)\propto\exp\biggl\{-\frac{1}{2B^{2}}\int_{0}^{\tau}dt\left[\left\{\dot{x}-\left(A\left(x\right)+u\left(x\right)\right)\right\}^{2}+B^{2}\frac{d\left(A+u\right)}{dx}\right]\biggr\}.
\end{equation}
By employing the Cameron-Martin formula, we have 
\begin{equation}
P_{u}\left(X|x_{0}\right)=\exp\left[-\frac{1}{2B^{2}}\int_{0}^{\tau}dt\left\{-2\dot{x}u\left(x\right)+\left(2A\left(x\right)+u\left(x\right)\right)u\left(x\right)+B^{2}\frac{\partial u\left(x\right)}{\partial x}\right\}\right]P_{0}\left(X|x_{0}\right).
\end{equation}
The transform, $u\left(x\right)=kB^{2}+B^{2}\partial\log\psi\left(x\right)/\partial x$, leads us to 
\begin{equation}
P_{u}\displaystyle \left(X|x_{0}\right)=\frac{\psi\left(x_{\tau}\right)}{\psi\left(x_{0}\right)}\exp\left[k\int_{0}^{\tau}dt\left\{\dot{x}-k\left(A\left(x\right)+\frac{kB^{2}}{2}\right)-\frac{\hat{L}_{k}^{\dagger}\psi\left(x\right)}{\psi\left(x\right)}\right\}\right]P_{0}\left(X|x_{0}\right),\label{add3}
\end{equation}
where $\hat{L}_{k}^{\dagger}$ is the differential operator, $\hat{L}_{k}^{\dagger}=\left(A\left(x\right)+kB^{2}\right)\partial/\partial x+\left(B^{2}/2\right)\partial^{2}/\partial x^{2}$.
Here. $\psi\left(x\right)$ is positive function and plays a role of the tuning function instead of $u\left(x\right)$. 
In order to find the optimal solution $P_{k}$ by tuning $u\left(x\right)$ (i.e. by tuning $\psi\left(x\right)$), a function $\psi_{*}\left(x\right)$  must exist such that $P_{k}=P_{u}$. 
Thus, we will prove existence of $\psi_{*}\left(x\right)$. 
By taking into account Eqs. (\ref{add2}) and (\ref{add3}), we can represent $P_{k}=P_{u}$ as 
\begin{equation}
\displaystyle \exp\left[k\int_{0}^{\tau}dt\left\{\dot{x}-\lambda_{0}\left(k\right)\right\}\right]P_{0}\left(X|x_{0}\right)=\frac{\psi_{*}\left(x_{\tau}\right)}{\psi_{*}\left(x_{0}\right)}\exp\left[k\int_{0}^{\tau}dt\left\{\dot{x}-k\left(A\left(x\right)+\frac{kB^{2}}{2}\right)-\frac{\hat{L}_{k}^{\dagger}\psi_{*}\left(x\right)}{\psi_{*}\left(x\right)}\right\}\right]P_{0}\left(X|x_{0}\right).\label{add4}
\end{equation}
Since the factor $\psi_{*}\left(x_{\tau}\right)/\psi_{*}\left(x_{0}\right)$ in right-hand side of Eq. (\ref{add4}) can be neglected for $\tau\rightarrow\infty$, we obtain
\begin{equation}
\displaystyle \dot{x}-\lambda_{0}\left(k\right)=\dot{x}-k\left(A\left(x\right)+\frac{kB^{2}}{2}\right)-\frac{\hat{L}_{k}^{\dagger}\psi_{*}\left(x\right)}{\psi_{*}\left(x\right)}.\label{add5}
\end{equation}
By simplifying Eq. (\ref{add5}), we attain 
\begin{equation}
\left\{\hat{L}_{k}^{\dagger}+k\left(A\left(x\right)+\frac{kB^{2}}{2}\right)\right\}\psi_{*}\left(x\right)=\lambda_{0}\left(k\right)\psi_{*}\left(x\right).
\end{equation}
From the Perron-Frobenius theory, we find that the differential operator, namely $\left\{\hat{L}_{k}^{\dagger}+k\left(A\left(x\right)+kB^{2}/2\right)\right\}$, have a positive eigenfunction [see Appendix B in Ref. \cite{23}]. 
Accrodingly, we have proved existence of $\psi_{*}\left(x\right)$.

\section{Derivation of Eq. (\ref{klm})}
We give the detailed calculation to derive Eq. (\ref{klm}). 
From Eqs. (\ref{pdo}) and (\ref{pdm}), we obtain 
\begin{eqnarray}
\displaystyle \frac{1}{\tau}\log\frac{P_{u}\left(X|x_{0}\right)}{P_{0}\left(X|x_{0}\right)}=\frac{1}{B^{2}}\frac{1}{\tau}\int_{0}^{\tau}dt\biggl\{u\left(x\right)\dot{x}-\left\{2A\left(x\right)+u\left(x\right)\right\}\frac{u\left(x\right)}{2}-\frac{B^{2}}{2}\frac{du\left(x\right)}{dx}\biggr\}.\label{b1}
\end{eqnarray}
The expectation of both sides in Eq. (\ref{b1}) leads us to 
\begin{eqnarray}
\displaystyle \left\langle\frac{1}{\tau}\log\frac{P_{u}\left(X|x_{0}\right)}{P_{0}\left(X|x_{0}\right)}\right\rangle_{u}=\mathbb{E}_{x_{0}}\biggl[\frac{1}{B^{2}}\frac{1}{\tau}\biggl\{\int_{x_{0}}^{x_{\tau}}u\left(x_{t}\right)\circ dx_{t}-\int_{0}^{\tau}\left[\left\{2A\left(x_{t}\right)+u\left(x_{t}\right)\right\}\frac{u\left(x_{t}\right)}{2}+\frac{B^{2}}{2}\frac{du\left(x_{t}\right)}{dx_{t}}\right]dt\biggr\}\biggr],\label{b2}
\end{eqnarray}
where $\mathbb{E}_{x_{0}}$ denotes the expectation of the Langevin process $x_{t}$ solving Eq. (\ref{mle}) under the initial condition $x_{0}$, and the symbol $\circ$ stands for the multiplication in the sense of the Stratonovich \cite{28,29}. 
By substituting Eq. (\ref{mle}) into Eq. (\ref{b2}) and changing the multiplication type from the Stratonovich sense to the Ito sense, we have
\begin{eqnarray}
\displaystyle \nonumber\left\langle\frac{1}{\tau}\log\frac{P_{u}\left(X|x_{0}\right)}{P_{0}\left(X|x_{0}\right)}\right\rangle_{u}&=&\mathbb{E}_{x_{0}}\biggl[\frac{1}{B^{2}}\frac{1}{\tau}\int_{0}^{\tau}\biggl\{u\left(x_{t}\right)\left\{A\left(x_{t}\right)+u\left(x_{t}\right)\right\}\\
\nonumber&&-\left\{2A\left(x_{t}\right)+u\left(x_{t}\right)\right\}\frac{u\left(x_{t}\right)}{2}-\frac{B^{2}}{2}\frac{du\left(x_{t}\right)}{dx_{t}}\biggr\}dt+\frac{1}{\tau}\int_{0}^{\tau}\frac{1}{B^{2}}u\left(x_{t}\right)\circ dW\biggr]\\
&=&\mathbb{E}_{x_{0}}\biggl[\frac{1}{\tau}\int_{0}^{\tau}\frac{u^{2}\left(x_{t}\right)}{2B^{2}}+\frac{1}{\tau}\int_{0}^{\tau}\frac{u\left(x_{t}\right)}{B^{2}}dW\biggr].
\end{eqnarray}
Calculating the expectation in the limit of $\tau\rightarrow\infty$, we finally find 
\begin{equation}
\displaystyle \left\langle\frac{1}{\tau}\log\frac{P_{u}\left(X|x_{0}\right)}{P_{0}\left(X|x_{0}\right)}\right\rangle_{u}=\int dx\rho_{\mathrm{s}\mathrm{t},u}\left(x\right)\frac{u^{2}\left(x\right)}{2B^{2}}.
\end{equation}



\begin{thebibliography}{00}

\bibitem{02}   Callen, H. B., Thermodynamics and an Introduction to Thermostatistics 2nd ed., Wiley, New York (1985).

\bibitem{03}   Dembo, A., and Zeitouni, O., Large Deviations Techniques and Applications 2nd ed., Springer (1998).

\bibitem{04}   Ellis, R. S., Entropy, Large Deviations, and Statistical mechanics, Springer (1985).

\bibitem{05}   Touchette, H., ``{\it The large deviation approach to statistical mechanics},'' Phys. Rep., {\bf 478}: 1-69 (2009). 

\bibitem{26}   G\"{a}rtner, J., ``{\it On large deviations from the invariant measure},'' Theory Probab. Appl., {\bf 22}: 24-39 (1977).

\bibitem{27}   Ellis, R. S., ``{\it Large deviations for a general class of random vecters},'' Ann. Probab., {\bf 12(1)}: 1-12 (1984).

\bibitem{28}   Gardiner, C. W., Handbook of Stochastic Methods 2nd ed., Springer (1985).

\bibitem{29}   Risken, H., The Fokker-Planck Equation 2nd ed., Springer (1989).

\bibitem{22}   Nemoto, T., and Sasa, S., ``{\it Variational formula for experimental determination of high-order correlations of current fluctuation in driven systems},'' Phys. Rev. E, {\bf 83}: 030105(R) (2011).

\bibitem{23}   Nemoto, T., and Sasa, S., ``{\it Thermodynamic formula for the cumulant generating function of time-averaged current},'' Phys. Rev. E, {\bf 84}: 061113 (2011).

\bibitem{24}   Donsker, M. D., and Varadhan, S. R. S., ``{\it Asymptotic evaluation of certain Markov process expectations for large time. I},'' Commum. Pure Appl. Math., {\bf 28}: 1-47 (1975).

\bibitem{99}   Sugiyama, Y., and Ohzeki, M., (in preparation).
\end{thebibliography}
\end{document}